\documentclass[11pt]{article}
\usepackage{graphicx}

\begin{document}

\def \p {\partial}
\def \dd {\psi_{u\bar dg}}
\def \ddp {\psi_{u\bar dgg}}
\def \pq {\psi_{u\bar d\bar uu}}
\def \jpsi {J/\psi}
\def \psip {\psi^\prime}
\def \to {\rightarrow}
\def\bfsig{\mbox{\boldmath$\sigma$}}
\def\DT{\mbox{\boldmath$\Delta_T $}}
\def\xit{\mbox{\boldmath$\xi_\perp $}}
\def \jpsi {J/\psi}
\def\bfej{\mbox{\boldmath$\varepsilon$}}
\def \t {\tilde}
\def\epn {\varepsilon}
\def \up {\uparrow}
\def \dn {\downarrow}
\def \da {\dagger}
\def \pn3 {\phi_{u\bar d g}}

\def \p4n {\phi_{u\bar d gg}}

\pagestyle{plain}
\vskip 10mm
\begin{center}
{\bf\Large Nonperturbative Corrections with Nonlocal Operators}\\
\vskip12pt
{\bf\Large to Lifetime Ratios of Beauty Hadrons  }\\
\vskip 10mm
J.P. Ma   \\
{\small {\it Institute of Theoretical Physics , Academia
Sinica, Beijing 100080, China \ \ \ }} \\
{\small {\it Department of Physics, Shandong University, Jinan Shandong 250100, China}}
\\
Z.G. Si \\
{\small {\it Department of Physics, Shandong University, Jinan Shandong 250100, China}}
\end{center}
\vskip 0.4 cm

\begin{abstract}
The motion of spectator quarks in decay of a beauty hadron is a nonperturbative
effect which can usually be neglected. We find that the motion in some decay channels,
which contribute total decay widths
of beauty hadrons, can not be neglected.
The contributions from these decay channels to decay widths are proportional
to certain averages of the squared inverse of the momentum carried by
a spectator quark. This fact results in that these contributions,
suppressed by $1/m_b^3$ formally,
are effectively suppressed by $1/m_b$. We find these contributions
can be factorized into products of perturbative coefficients and
nonperturbative parameters. We calculate these coefficients and define
these nonperturbative parameters in terms of HQET matrix elements.
Since these parameters are unknown, we are unable to give numerical predictions
in detail. But with a simple model it can be shown that these contributions
can be large.
\end{abstract}
\par\vfil\eject

Although the heavy quark effective theory(HQET) is very successful
to predict various properties of hadrons containing one heavy
quark\cite{HQET}, but it is still difficult to predict the experimentally
observed ratio of decay widths of $B$-meson and $b$-flavored
baryon. The experimental results for lifetime ratios of beauty hadrons
are\cite{exp}:
\begin{equation}
\frac{\tau (B^+)}{\tau (B_d)} = 1.074\pm 0.014,
\ \ \ \
\frac{\tau(B_s)}{\tau (B_d)} = 0.948\pm 0.038,
\ \ \ \
\frac{\tau(\Lambda_b)}{\tau(B_d)} = 0.796\pm 0.052.
\end{equation}
In HQET a systematic expansion in $1/m_b$ is employed to give
theoretical predictions. At the leading order, the decay width of
a beauty hadron equals the decay width of a free $b$-quark,
hence the above ratio should be one, if higher-order correction
is neglected. It is clearly that from Eq.(1) the ratio with $\Lambda_b$
deviates from one substantially.
\par
The correction to the leading-order result starts at order of $m_b^{-2}$. The decay
width of a $b$-flavored hadron, denoted as $H_b$,  takes in
general the form
\begin{equation}
 \Gamma (H_b) = \Gamma (b) \left ( 1 +\sum_{n=2} c_n \frac {\langle H_b \vert {\cal O}_n \vert H_b\rangle}{m_b^n}
               \right ),
\end{equation}
where $c_n$ is a perturbative coefficient, ${\cal O}_n$'s are operators
of HQET, whose matrix element represents nonperturbative effects
in the decay. So far only contributions from local operators are considered.
It should be noted that the correction at order
of $m_b^{-1}$ does not exist. It is expected
that spectator effects will results in the difference
between life times of $b$-flavored hadrons.
Spectator effects appear at order of $m_b^{-3}$.
These effects have been studied in \cite{BUV,Neubert} at
tree-level of QCD. Next-to-leading order corrections
have been studied\cite{BBGLN,CFLM,FLMT}.
In these studies the nonperturbative effects at order of $m_b^{-3}$
are parameterized with matrix elements of four quark operators.
These matrix elements have been studied with lattice QCD\cite{Lattice}
or with sum rule techniques(e.g., see \cite{SumRule}).
With obtained values of matrix elements and effects at next-to-leading
order the ratio becomes\cite{FLMT}:
\begin{equation}
\frac{\tau (B^+)}{\tau (B_d)} = 1.06\pm 0.02,
\ \ \ \
\frac{\tau(B_s)}{\tau (B_d)} = 1.00\pm 0.01,
\ \ \ \
\frac{\tau(\Lambda_b)}{\tau(B_d)} = 0.90\pm 0.05.
\end{equation}
This prediction is closer to experiment, but there is still
a discrepancy at order of $10\%$ for $\Lambda_b$. Other attempts to explain
the ratio in Eq.(1) can be found in \cite{AMPR}.
It is interesting
to note that the effects at order of $m_b^{-3}$ studied before consist
of local operators of
four quark fields, in which two fields are fields of HQET for $b$-quark,
while other two are for light quarks. These light quarks
can be either as a part of $H_b$ as a bound state, or they generate
nonperturbatively soft decay products.
Since the operators are local,
the light quarks represented by the two quark fields carry
zero momenta in the decay, i.e., the motion of spectator quarks
is neglected.
\par
The formula in Eq.(2) is based on a factorization in HQET, in which
nonperturbative effects are factorized from perturbative effects.
In general, one can expect that similar factorization formula
can be obtained for inclusive productions of $H_b$, where
nonperturbative effects related to $H_b$ are parameterized
by matrix elements of local operators, in corresponding
to those matrix elements of $\langle H_b \vert {\cal O}_n \vert H_b\rangle$.
At the leading order of $m_b^{-1}$ the production rate of $H_b$
can be factorized as a product of the production rate of a free
$b$-quark with a matrix element of HQET. The matrix element
can be interpreted as the probability for the transition of the
$b$-quark into $H_b$.
Inclusive productions based on such a factorization have been studied,
predictions at leading order of $m_b^{-1}$ have been made
for production at $e^+e^-$ colliders and for polarization
of heavy vector meson\cite{Ma1,Ma2}. In these cases a good
agreement with experiment were found.
\par
Recently, such a factorization was employed to explain the
asymmetry between production rates of $D^+$ and of $D^-$ in
their inclusive productions, and also asymmetries
for other heavy flavored hadrons\cite{B1,B2,B3,ZMS}.
In these works contributions from quark recombination
to production rates are studied, in which a heavy flavored hadron
like $H_b$ is produced by combining a $b$-quark with a light
antiquark $\bar q$. Including these contributions
the asymmetry can be explained\cite{B1,B2,B3}.
The nonperturbative effect of the recombination
can be represented as matrix elements of four quark fields in the
production rate. But it is found that the momentum of $\bar q$
can not be taken as zero, because it will generate a type of
infrared singularities in the production amplitude of $b\bar q$.
This type of singularities can be regularized
by noting the fact that the light antiquark inside $H_b$
carries a small momentum at order of $\Lambda_{QCD}$
and its effect is nonperturbative.
One needs new matrix elements beside these
matrix elements of {\it local} operators, corresponding
to those in Eq.(2), to incorporate this nonperturbative effect.
These new matrix elements are found to be {\it nonlocal}
matrix elements, whose definitions are given in \cite{ZMS}.
It turns out that the contributions to the production rates
due to quark recombination is proportional some averages
of the inverse of momenta carried by the light antiquark.
This results in that the contributions are significant
and the large asymmetry observed in experiment
can be explained in this way. The effect in quark recombination
in production $c$-flavored baryon\cite{BMJM} and in
production $b$-flavored jet in $Z^0$-decay\cite{Jia}
has also been studied.
\par
In this letter we point out similar contributions also exist as
corrections to lifetime ratios. These contributions formally are suppressed
by $m_b^{-3}$, but they are proportional to the square of the
inverse of the momentum carried by a light antiquark in $H_b$ or
by a light antiquark which generates soft decay products
nonperturbatively. Therefore these corrections are only suppressed
by $m_b^{-1}$ effectively.
Similar situation also appears in the decay $B\to \gamma\ell\nu$,
where the decay amplitude is proportional to a certain average
of the inverse of the momentum carried by the light quark inside
the $B$ meson. This decay mode has been studied in detail in \cite{blv,blv1}.
In this letter we will study this type of contributions to lifetimes of
beauty hadrons.
\par
The effective weak Hamiltonian for the
decay of $H_b$ at the scale $\mu=m_b$ is:
\begin{eqnarray}
H_{eff} &=& \frac{G_F}{\sqrt{2}} V_{cb}\sum_{q=d,s} \Big \{
 c_1 (m_b) \big [ V^*_{uq} \bar q_L \gamma^\mu u_L \bar c_L
 \gamma_\mu b_L + V^*_{cq} \bar q_L\gamma^\mu c_L \bar c_L
 \gamma_\mu b_L \big ]
 \nonumber\\
   && \ \ \ \ +c_2(m_b) \big [ V^*_{uq} \bar c_L \gamma^\mu u_L \bar q_L
   \gamma_\mu b_L + V^*_{cq} \bar c_L \gamma^\mu c_L \bar q_L
   \gamma_\mu b_L \big ] \Big\},
\end{eqnarray}
where we neglected the suppressed transition $b\to u$ and
$q_L=\frac{1}{2}(1-\gamma_5)q$. $c_1$ and $c_2$ are Wilson
coefficients. The above mentioned contributions come from these
decays at the tree-level:
\begin{eqnarray}
H_b(P) & \to &   c(p_1)+ q(p_2) +G(k) +X,
\nonumber \\
H_b(P) & \to & c(p_1) + \bar u (p_2) +G(k) +X,
\nonumber \\
H_b(P) &\to & c(p_1) +\bar c (p_2)  +G(k) + X.
\end{eqnarray}
where momenta are given in brackets. We will not consider
a $c$-quark as a spectator, hence the decay into a lepton pair
with a gluon and other unobserved states $X$ will not lead to
contributions we consider.
In general the gluon can have
any possible momentum. If the momentum is large, one can use
perturbative QCD. If the momentum is small, the contributions can
have an infrared singularity, reflecting the fact that the gluon
can not be taken as a perturbative gluon. However, it turns out
that the contributions we are interesting in are free of infrared
singularities. This will be discussed in detail. We will give
some detail for calculation the contribution from the process
$H_b \to c+q+G+X$.
\par
With the effective Hamiltonian the decay
amplitude can be written as:
\begin{equation}
{\cal T} = \int \frac{d^4 q_1 } {(2\pi)^4}  A_{ij}(q_1,q_2)
\int d^4 x_1e^{iq_1\cdot x_1} \langle X \vert u_i (x_1) b_j (0) \vert H_b \rangle ,
\end{equation}
where $i,j$ stand for spin- and color indices. $A_{ij}$ is the scattering
amplitude for $ u (q_1) + b(q_2) \to c(p_1) +q (p_2) +G(k)$
in which the quarks in the initial state are off-shell in general and
$q_1+q_2 =p_1+p_2+k$.
With the translational symmetry the contribution to the decay width can be written:
\begin{eqnarray}
\delta\Gamma_{cqg} &=& \int d\Gamma_{cqg}\int \frac{d^4 q_1}{(2\pi)^4}
          \frac{d^4 q_3}{(2\pi)^4}
          A_{ij}(q_1,q_2) \left ( \gamma^0 A^\dagger(q_3,q_4) \gamma^0\right)_{lk}
  \nonumber\\
 && \cdot (-1) \int d^4 x_1 d^4 x_3 d^4 x_4 ~
   e^{ iq_1\cdot x_1 -iq_3\cdot x_3 -i x_4\cdot q_4 }
  \langle H_b \vert \bar b_l (x_4) u_i(x_1) \bar u_k(x_3) b_j(0)
    \vert H_b \rangle ,
\end{eqnarray}
with $q_4 = p_1+p_2+k-q_3$. In the above we exchanged
the order of $u$-quark fields and this gives the $-$ sign
in the second line. The integration measure for the phase space
of three particles $cqg$ is denoted as $d\Gamma_{cqg}$.
We use nonrelativistic normalization
for the state $H_b$ and $b$-quark. The above contribution can
be illustrated by Fig.1., where one of 16 diagrams is shown
explicitly.

\begin{figure}[hbt]
\centering
\includegraphics[width=8cm]{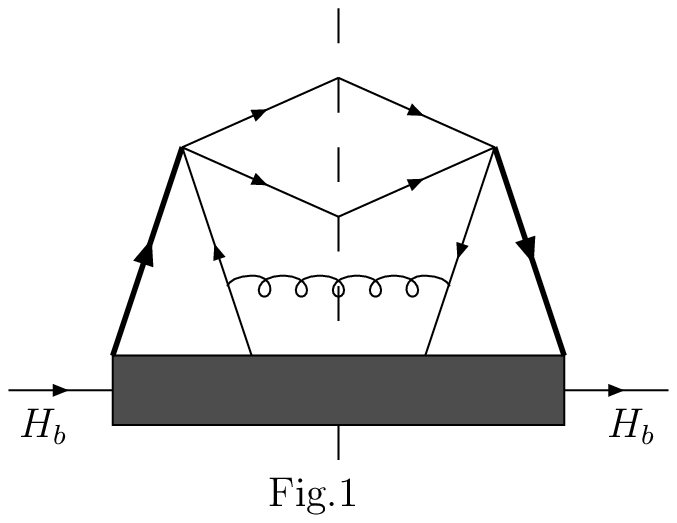}
\caption{Diagrams for the contributions to the decay width.
Other diagrams are obtained by changing attachments of the gluon. The thick
line is for the $b$-quark. The
broken line is the cut, the black box represents nonperturbative effect
in the decay, its expression is given in Eq.(9).}
\label{Feynman-dg1}
\end{figure}
\par
\par

\par
For $b$-quark field $b(x)$ one can use the $m_b^{-1}$ expansion:
\begin{equation}
b(x) =e^{-im_b v\cdot x } \left ( h(x) +\cdots \right  ), \ \ \
\bar b(x) = e^{+im_b v \cdot x} \left ( \bar h(x) +\cdots \right  ),
\end{equation}
where $v$ the four velocity of $H_b$.
The $\cdots$ represent
higher orders in $m_b^{-1}$, which can be neglected in this letter.
With $v$ any vector $B^\mu$ can be
decomposed as $B^\mu =v\cdot B v^\mu + B^\mu_\perp$ with $B_\perp\cdot v=0$.
Taking the leading term the Fourier transformed matrix element becomes
\begin{eqnarray}
 && \int d^4 x_1 d^4 x_3 d^4 x_4 ~
   e^{ iq_1\cdot x_1 -iq_3\cdot x_3 -i x_4\cdot q_4 }
  \langle H_b \vert \bar b_l (x_4) u_i(x_1) \bar u_k(x_3) b_j(0)
    \vert H_b \rangle
  \nonumber \\
  && \ \ \   =\int d^4 x_1 d^4 x_3 d^4 x_4 ~
   e^{ iq_1\cdot x_1 -iq_3\cdot x_3 -i x_4\cdot (q_4-m_b v) }
  \langle H_b \vert \bar h_{l} (x_4) u_i(x_1) \bar u_k(x_3) h_{j}(0)
    \vert H_b \rangle + \cdots .
\end{eqnarray}
Since we extract the large momentum $m_b v$ by using the expansion in
Eq.(8), the space-time dependence of the matrix element in the second line
in Eq.(10) is controlled by the soft scale $\Lambda_{QCD}$.
The $x$-dependence of $h$ fields can be safely neglected.
If we can neglect the $x$-dependence of the light quark fields, then the contribution
will proportional to matrix elements of {\it local} four-quark
operators, which appear at order of $m_b^{-3}$ in Eq.(2).
This implies that the light quark
$\bar q$ will carry zero momentum. But the amplitude $A$ with
the zero momentum is divergent or proportional to the inverse of the light quark mass
$m_u$ if we do not neglect this mass.
This divergence represents
some new nonperturbative effects which can not be represented
by local four quark matrix elements.
To study the divergence, we introduce
a light-cone coordinate system in which the two light-cone vectors
are $l$ and $n$ respectively and $l\cdot n=1$.
In the light-cone coordinate system the emitted gluon has the momentum
$k^\mu =\sqrt{2}k^0 l^\mu$.
The expansion in $q_1$ reads:
\begin{equation}
A_{ij}(q_1, m_bv)
=-\frac{m_b}{q_1\cdot l} T_{ij} +\cdots,
\end{equation}
where $\cdots$ stand for higher order terms of $q_1$ and we set $q_2=m_b v$.
$T_{ij}$ reads:
\begin{eqnarray}
T_{ij} &=&  i \frac{g_s G_F}{\sqrt{2}m_b } V_{cb}V_{uq}^*
  \Big \{ c_1  \big [ \bar u(p_1) \gamma^\mu (1-\gamma_5) \big ]_j
                  \big [ \bar u(p_2) T^a \gamma_\mu (1-\gamma_5)
                   \gamma\cdot\varepsilon^*(k) \gamma\cdot l \big ]_i
        \nonumber\\
        && \ \ \ \ \ \ \ \
           -  c_2  \big [ \bar u(p_2) \gamma^\mu (1-\gamma_5) \big ]_j
                  \big [ \bar u(p_1) T^a \gamma_\mu (1-\gamma_5)
                   \gamma\cdot \varepsilon ^* (k) \gamma\cdot l \big ]_i \Big \}
  \nonumber \\
  & =& \left [ T' \gamma\cdot l \right ]_{ji}
\end{eqnarray}
where $\varepsilon^* (k)$ is the polarization vector of the gluon and
we used $k\cdot \varepsilon^* (k)=0$.
$T_{ij}$ only gets contribution from Fig.1., it does not depend
on $q_1$. It should be noted that the matrix $T$ can always be
written in the form as in the last line of Eq.(10) for our  processes in Eq.(5).
Keep only the leading term for $A_{ij}$, some integrations in Eq.(7)
can be performed. We obtain:
\begin{eqnarray}
\delta\Gamma_{cqg} &=& \int d\Gamma_{cqg}
         \int \frac{d\eta_1}{2\pi}
          \frac{d\eta_3}{2\pi}(2\pi)^4\delta^4(m_b v-p_1-p_2-k)
        \frac{1}{2\eta_1\eta_3}  T'_{ji} \left ( \gamma^0 T^{\prime\dagger} \gamma^0\right)_{kl}
  \nonumber\\
 && \cdot (-m_b^2) \int d \omega_1  d\omega_3 ~
   e^{ i\eta_1\omega_1m_b -i \eta_3 \omega_3m_b  }
  \langle H_b \vert \bar h_{l} (0) [\gamma^-u]_i(\omega_1 l )
  [\bar u \gamma^-]_k(\omega_3 l ) h_{j}(0)
    \vert H_b \rangle.
\end{eqnarray}
where we have used $q_i\cdot l = \eta_i m_b$ for $i=1,3$ and moved $\gamma^-=\gamma\cdot l$.
into the matrix element.
It is interesting to note that the $T_{ij}$ is finite when the gluon
carries null momentum. This indicates that the decay width will
be free from infrared singularities when we perform the phase-space
integration. If one keeps the next-to-leading order in $q_1$ or $q_3$,
infrared singularities
will appear, but these singularities may be cancelled by
some virtual corrections partly and absorbed into four quark matrix elements.
The contribution
at this order will be suppressed by $(M_{H_b}-m_b)/m_b$ in comparison
with that from the leading order.
In our approximation the decay width will be proportional
to the integral of
the Fourier transformed matrix element.
The Fourier transformed matrix element can be parameterized as:
\begin{eqnarray}
&&\frac{1}{m_b}\int \frac{d\eta_1}{\eta_1} \frac{d\eta_3}{\eta_3}
   \int  \frac{d\omega_1}{2\pi}  \frac {d\omega_3}{2\pi} ~
   e^{ i\eta_1\omega_1m_b -i \eta_3 \omega_3 m_b  }
  \langle H_b \vert \bar h_{l} (0) [\gamma^-u]_i(\omega_1 l )
  [\bar u\gamma^-] _k(\omega_3 l ) h_{j}(0)
    \vert H_b \rangle
       \nonumber\\
&& =
    \frac{1}{3} \big [ \left ( P_v\right )_{jk}
   \left ( P_v\right )_{il}
   {\cal S}_{H_b}^{(u,1)}
   - \left (  \gamma_5 P_v \right )_{jk}
   \left ( P_v\gamma_5\right )_{il}
    {\cal P}_{H_b}^{(u,1)}
 - \left (  \gamma_T^\mu P_v\right )_{jk}
   \left ( P_v\gamma_{T\mu}\right )_{il}
     {\cal V}_{H_b}^{(u,1)}
\nonumber\\
  && - \left (  \gamma_T^\mu\gamma_5 P_v\right )_{jk}
   \left ( P_v \gamma_{T\mu}\gamma_5\right )_{il}
        {\cal A}_{H_b}^{(u,1) } \big ]
   + \frac{1}{2} \big [ \left ( P_v T^a \right  )_{jk}
   \left ( P_v    T^a \right   )_{il}
   {\cal S}_{H_b}^{(u,8)}
   - \left (  \gamma_5 P_v T^a \right )_{jk}
   \left ( P_v\gamma_5T^a \right )_{il}
    {\cal P}_{H_b}^{(u,8)}
    \nonumber\\
 &&  - \left (  \gamma_T^\mu P_v T^a \right )_{jk}
   \left ( P_v\gamma_{T\mu}T^a \right )_{il}
     {\cal V}_{H_b}^{(u,8)}
   - \left (  \gamma_T^\mu\gamma_5 T^a P_v\right )_{jk}
   \left ( P_v \gamma_{T\mu}\gamma_5T^a \right )_{il}
        {\cal A}_{H_b}^{(u,8)} \big ],
\end{eqnarray}
with
\begin{equation}
\gamma^\mu_T=\gamma^\mu -v\cdot\gamma v^\mu, \ \ \ \
              P_v =\frac{1+\gamma\cdot v}{2}
\end{equation}
The above matrix elements are defined in the rest frame of $H_b$.
The eight coefficients ${\cal S}_{H_b}^{(u,1)},\cdots$ in Eq.(13) are dimensionless,
their values depend on $H_b$.
These coefficients multiplied with the factor $\sqrt{2v^0}$ are Lorentz covariant
because we take nonrelativistic normalization for the state.
We work in the light-cone gauge $l\cdot G =0$. In other gauges
gauge links should be added into the above matrix element between
quark fields to make it gauge invariant.
If we replace in the integral the factor $(\eta_1\eta_3)^{-1}$ with a constant,
the eight coefficients will be at order of $\Lambda_{QCD}^3/m_b^3$. Taking
this fact into account, the coefficients are at order of
$m_b^2/(M_{H_b}-m_b)^2  \cdot \Lambda_{QCD}^3/m_b^3 \sim  \Lambda_{QCD}/m_b$.
Therefore, the contribution to the total decay width
is effectively suppressed by $\Lambda_{QCD}/m_b$.
It is now straightforward to obtain:
\begin{eqnarray}
\delta\Gamma_{cqg} &=& -64 g_s^2 G_F^2 \vert V_{cb} V^*_{uq} \vert ^2 m_b
   \cdot \int d\Gamma_{cqg} (2\pi)^4 \delta^4(m_bv-p_1-p_2-k) p_1\cdot p_2
 \nonumber\\
   && \cdot  \Big \{ (c_1^2+c_2^2) \left [ \frac{4}{3} \left (
        f_1^{u/H_b}+f_2^{u/H_b}
           \right ) +\frac{16}{6}
           \left ( f_3^{u/H_b}+f_4^{u/H_b}
           \right ) \right ]
   \nonumber\\
    && \        +2 c_1c_2\left (f_3^{u/H_b}+f_4^{u/H_b}
           \right ) \Big \}.
\end{eqnarray}
The parameters ${\cal W}_{S-P}^{(u,i)},\cdots$ are defined as:
\begin{eqnarray}
f_1^{u/H_b} &=& {\cal S}_{H_b}^{(u,1)}+ {\cal P}_{H_b}^{(u,1)}, \ \ \ \
f_2^{u/H_b} ={\cal V}_{H_b}^{(u,1)} + {\cal A}_{H_b}^{(u,1)},
\nonumber\\
f_3^{u/H_b} &=& {\cal S}_{H_b}^{(u,8)}+ {\cal P}_{H_b}^{(u,8)}, \ \ \ \
f_4^{u/H_b} ={\cal V}_{H_b}^{(u,8)} + {\cal A}_{H_b}^{(u,8)},
\end{eqnarray}
Performing the phase-space integration we obtain:
\begin{eqnarray}
\delta \Gamma_{cqg}&=&-32\pi \alpha_s \Gamma_0 |V_{uq}^*|^2
\Big\{1-6z+3z^2+2z^3-6z^2\ln(z)\Big\}
\nonumber\\
&& \cdot \Big\{ \frac{4}{3} (c_1^2+c_2^2)[f_1^{u/H_b}+f_2^{u/H_b}]+
[\frac{8}{3} (c_1^2+c_2^2)+2c_1 c_2) [f_3^{u/H_b}+f_3^{u/H_b}] \Big\}  ,
\end{eqnarray}
where $z=\frac{m_c^2}{m_b^2}$ and
\begin{equation}
\Gamma_0 = \frac{G_F^2 m_b^5}{192\pi^3} \vert V_{cb} \vert^2.
\end{equation}
\par
Similarly, one can work out the contributions of
other two processes. To present our results we introduce:
\begin{eqnarray}
 F_{cug} & =& \frac{1}{3}\big [ 2 - 9z + 18z^2 - 11z^3 + 6z^3\ln (z)\big ],
\nonumber\\
\tilde F_{cug} &=& \frac{1}{6}
\big [ 11 - 54z + 36z^2 - 2z^3 + 9z^4 - 12z^2(3 + z)\ln (z) \big ],
\nonumber\\
F_{ccg}&=&\frac{2\beta}{3}\Big[1-7z+6z^2]-2 (5+2z^3)
\ln\Big[1-3z-\beta(1-z)\Big]
\nonumber \\
&&+4z^3 \ln[z(1+\beta)]+10\ln\Big[\frac{4z^2(1-\beta)}{(1+\beta)^2}\Big],
\nonumber \\
\tilde{F}_{ccg}&=&\frac{\beta}{6}\Big[11-86z-6z^2+108 z^3\Big]
-10\ln\Big[\frac{4z^2(1-\beta)}{(1+\beta)^2}\Big]+2 \Big(5-3z^2+2z^3 \nonumber
\\
&&-9z^4\Big)\ln\Big[1-3z-\beta
(1-z)\Big]+2z^2\Big(3-2z+9z^2\Big)\ln\Big[z(1+\beta)\Big],
\end{eqnarray}
with $\beta=\sqrt{1-4z}$.
The processes $H_b(P)\rightarrow c(p_1)+\bar{u}(p_2)+G(k)+X$ gives the contribution:
\begin{eqnarray}
\delta \Gamma_{cug}&=&32 \pi \alpha_s \Gamma_0 |V_{uq}^*|^2 \Big\{
\frac{2}{3} c_1^2 \Big[f_1^{q/H_b}\, F_{cug}
+f_2^{q/H_b}\tilde F_{cug}\Big] \nonumber \\
&&+\Big[\frac{4}{3}c_1^2+\frac{1}{2} \Big(c_2^2+2c_1c_2\Big)\Big]
  \Big[f_3^{q/H_b}\, F_{cug}
+f_4^{q/H_b}\tilde F_{cug}\Big]\Big\},
\end{eqnarray}
Replacing the $\bar u$ quark with a $\bar c$ quark, we obtain
the contribution $\delta \Gamma_{c\bar c g}$ from the process
$H_b(P)\rightarrow c(p_1)+\bar{c}(p_2)+G(k)+X$.
\par
We use the parameters:
\begin{equation}
m_b=4.8{\rm GeV}, \ \ \ \  z=0.085.
\end{equation}
Correspondingly we have
$\alpha_s =0.18,~c_1=1.105$ and $c_2=-0.245$. The lifetime ratio
can be predicted by using experimental values as:
\begin{equation}
\frac{\tau(H_b)}{\tau (B_d)}-1 =\frac{\Gamma(\bar B^0)}{\Gamma (H_b)} -1
   =\left ( \frac{\Gamma(\bar B^0)}{\Gamma (H_b)}\right )_{exp}
      \cdot \frac{\Gamma_0}{\Gamma_{exp} (\bar B^0)} \cdot
      \left ( \frac{ \Gamma(\bar B^0) -\Gamma (H_b)}{\Gamma_0} \right )_{theory}.
\end{equation}
For our numerical estimation we only take effects of
valence quarks into account and use isospin symmetry.
We obtain:
\begin{eqnarray}
\frac{\tau (B^-)}{\tau (B^0)}&=&
1+5.06 f_1^{u/B^-}+7.82 f_2^{u/B^-}
                        +8.55 f_3^{u/B^-}+13.26 f_4^{u/B^-},
\nonumber \\
\frac{\tau (B_s)}{\tau (B^0)}&=&
1-0.94 f_1^{s/B_s}+1.50 f_1^{u/B^-}
          -2.55 f_2^{s/B_s} +4.07 f_2^{u/B^-}-1.59  f_3^{s/B_s}
\nonumber
\\
&&          +2.56  f_3^{u/B^-} -4.34 f_4^{s/B_s}+6.94 f_4^{u/B^-},
\nonumber\\
\frac{\tau (\Lambda_b)}{\tau (B^0)}&=&1+1.36 f_1^{d/\Lambda_b}+1.20 f_1^{u/B^-}
 -0.69 f_2^{d/\Lambda_b}+3.25 f_2^{u/B^-}+2.27 f_3^{d/\Lambda_b}
\nonumber
\\
&&     +2.04 f_3^{u/B^-} -1.23 f_4^{d/\Lambda_b}+5.53 f_4^{u/B^-}.
\end{eqnarray}
At moment no information for these nonperturbative parameters
is available. This prevents us to give numerical predictions.
If we use the approximation of vacuum saturation, one may
estimate the correction to lifetimes of beauty mesons.
We use this approximation below to give some numerical
predictions for mesons.
\par
If one uses the vacuum saturation approximation, the nonperturbative
parameters are either zero or can be expressed with the light-cone wave
function $\phi_+$ of $B$-meson, which has been studied extensively(See e.g.,
\cite{BW00,BW1,BW2,BW3}.
The light cone wave function is defined in the light-cone gauge as\cite{BW3}:
\begin{equation}
\langle 0 \vert \bar q (zl)\gamma\cdot l \Gamma h(0) \vert \bar B(v) \rangle
= -\frac{i}{2} F_B {\rm Tr}\left [ \gamma_5 \gamma\cdot l \Gamma
  P_v \right ]
  \phi_+(z),
\end{equation}
where $F_B$ is the decay constant in HQET. At tree-level it is related
to the decay constant $f_B$ of full QCD via $F_B=f_B \sqrt{M_B}$.
$\Gamma$ is an arbitrary Dirac matrix. Under the approximation of
vacuum saturation, one can show with Eq.(24) that all nonperturbative
parameters in Eq.(13) are zero except ${\cal P}_{B}^{(q,1)}$.
Hence only $f_1^{q/B}$ is not zero and it is given by:
\begin{equation}
f_1^{q/B} = \frac{f_B^2 M_B}{4m_b}\cdot  \frac{1}{\lambda_B^2}, \ \ \
    \frac{1}{\lambda_B} =\left  \vert
\int \frac{d k}{k} \phi_+ (k)\right \vert,
\end{equation}
where $\phi_+(k)$ is the Fourier transformed function of $\phi_+(z)$.
The parameter $\lambda_B$ is at order of $\lambda_{QCD}$. It is estimated
to be in the range $0.35\sim 0.6$MeV\cite{blv1,BW3,lb1,lb2}. With this
we obtain the following numerical results in the approximation of vacuum
saturation:
\begin{eqnarray}
\frac{\tau (B^-)}{\tau (B^0)}&=& 1+ (0.15\sim 0.45),
\nonumber \\
\frac{\tau (B_s)}{\tau (B^0)}&=& 1 + (0.017\sim 0.049)
\end{eqnarray}
where we have used $SU(3)$ isospin symmetry. Comparing experimental results,
the predicted ratio of $B^-$ is too large. It should be noted that
the approximation of vacuum saturation is not a well-established
approximation, it serves only for a rough estimation. A detailed study
of these nonperturbative parameters is required. If one assumes
that the nonperturbative parameters of $\Lambda_b$ is at the same order
of $f_1^{q/B}$ determined as above, one can also have a large correction
to the ratio of $\Lambda_b$.
\par
To summarize: In this letter we have studied corrections of spectator quarks
to lifetime ratios of beauty hadrons. Formally, these corrections
are at order of $m_b^{-3}$.
We find some decay channels which
lead to that these corrections are proportional
to certain averages of the squared inverse of the momentum carried by
a spectator quark. Hence these corrections are effectively at order
of $m_b^{-1}$. The corrections are calculated at tree-level and
nonperturbative effects are parameterized with nonlocal operators.
Since the nonperturbative parameters are unknown,
we are unable to give numerical predictions in detail.
With a simple model one can estimate these parameters and the obtained
ratio of $B^-$ is too large in comparison with experiment.
However, estimations of these parameters in order of magnitude
indicate that these corrections may be large enough to accommodate experimental
results. A detail study
of these nonperturbative parameters is required to have
detailed predictions.
\par
\vskip20pt
{\bf Acknowledgements}
\par
This work of is supported by National Nature
Science Foundation of P. R. China.
\vfil\eject

\end{document}